\documentclass[aps,pra,twocolumn,groupedaddress]{revtex4-1}

\usepackage{amsmath}
\usepackage{amsthm}

\usepackage{bbm,graphicx,bm,hyperref}   
\usepackage[pdftex]{color}

\def\R{R\'enyi }
\def\ii{{\rm i}}

\def\sz#1{\sigma^{\rm z}_{#1}}
\def\tx#1{\tau^{\rm x}_{#1}}
\def\ty#1{\tau^{\rm y}_{#1}}

\def\ket#1{{| #1 \rangle}}

\def\tit#1{{\em #1},}

\newcommand{\new}[1]{{\color{black} #1}}

\begin{document}

\title{Response to ``Entanglement growth in diffusive systems with large spin''}

\author{Marko \v Znidari\v c}
\affiliation{Department of Physics, Faculty of Mathematics and Physics, University of Ljubljana, 1000 Ljubljana, Slovenia}

\date{\today}

\begin{abstract}
Reply to a comment by T.~Rakovszky, F.~Pollmann, and C.~W~von Keyserlingk~\cite{comment}.
\end{abstract}


\maketitle

The question discussed in Refs.~\cite{my,comment} is about the growth of the 2nd \R entropy $S_2$ at long times in systems with a diffusive degree of freedom (DOF). One in particular wants to distinguish between the diffusive $S_2 \sim \sqrt{t}$ (found for any diffusive system with the local Hilbert space dimension $q=2$ and spatial dimension $d=1$~\cite{prl19}) and ballistic $S_2 \sim t$ growth (generic~\cite{my} for $q>2$). Contested~\cite{comment} is the case of single diffusive DOF (like charge, density, spin,...) in systems with $q>2$.

The correct thermodynamic limit (TDL) is to first let the system size to infinity, $L \to \infty$, only then $t \to \infty$. In order to be able to unanonimously distinguish different asymptotic powers in $S_2 \sim t^\alpha$ one must have an infinite range of possible values that $S_2$ can take (at least in principle~\cite{foot2}), which necessarily implies that the size of the subsystem $A$ must be infinite. Therefore, (a) we study situations where the size of the subsystem $A$ diverges in the TDL. \new{We will also use guiding principles of Science}~\cite{EB}: (b) description of the physical world, (c) unbiased observations (present all evidence, not just the one in favor of a chosen narrative), (d) experimentation (verifiability).

\new{Let me first summarize Ref.~\cite{my} (Ref.~\cite{comment} in the abstract incorrectly says that~\cite{my} claims the diffusive growth of $S_2$ appears only in $d=1$ and $q=2$).} Ref.~\cite{my} presents arguments and shows a compelling numerical evidence that for $q>2$ one will in general observe ballistic growth of $S_2$, however, it does not exclude diffusive growth for $q>2$ and $d=1$. It presents a number of examples ($q=3$ and $q=4$) that in fact do display diffusive growth. \new{While presenting an additional diffusive system~\cite{comment} is useful in delineating cases with $q>2$ where diffusive growth nevertheless does occur, it does not invalidate the main message of Ref.~\cite{my}.}

The main point of Ref.~\cite{comment} seems to be (mentioned e.g. in the abstract and the 1st paragraph) that there are ``more'' diffusive cases in $q>2$. \new{We all agree that there are diffusive cases, but to be able to say which are more one has to count and \cite{comment} does not specify how they count Hamiltonians. I will instead focus on well defined questions.}

That being said, in Ref.~\cite{my} I do use the word ``generic'', so let me explain what it means. It means generic in the sense of point (b) -- describing nature. Elementary particles all have small internal dimension and so lattice models with large $q$ will be typically obtained by a direct product of such elementary DOF. Large $q$ will be a consequence of having multiple (interacting) DOF at a single lattice site (e.g., spin, charge, multiple fermion species, bosons,...)~\cite{foot3}, each of which can or can not be diffusive. For instance, a canonical example is the Hubbard model with $q=4$ due to spin-up and spin-down fermions. If in such models only a single 2-level DOF is conserved and diffusive, and there are no additional constrains, one will observe $S_2 \sim t$ -- and that is the main message of Ref.~\cite{my} (which is not in conflict with~\cite{prl19,huang20}). Simply put, a single diffusive DOF is in itself a weak constraint in a large Hilbert space.

\begin{figure}[t!]
\centerline{\includegraphics[width=1.0\columnwidth]{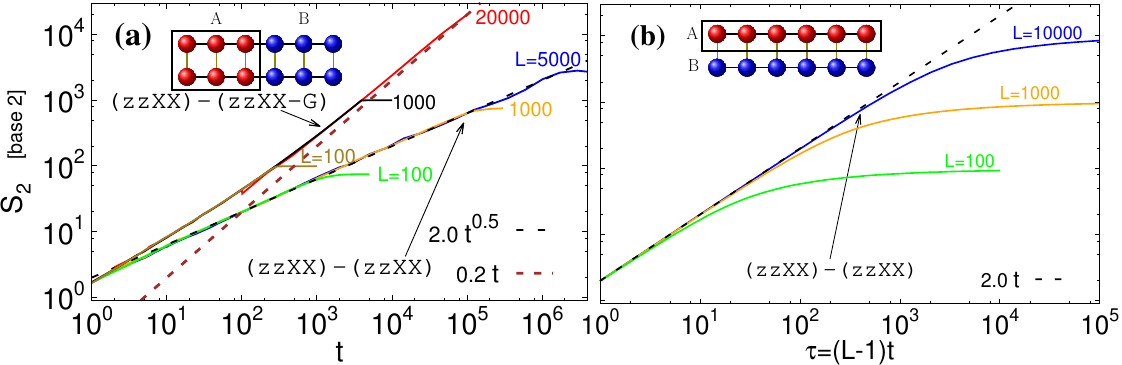}}
\caption{(a) Ballistic growth of $S_2$ in a Floquet ladder model with a single diffusive DOF (zzXX-(zzXX-G)) vs. diffusive for two diffusive DOF (zzXX-zzXX). (b) Even if spin on both legs is diffusive one can get the ballistic growth for the shown bipartition to A and B. The initial state is $(\ket{0}+\ket{1})^{\otimes 2L}$.}
\label{fig:plotzzXX}
\end{figure}

Ref.~\cite{my} presents Floquet Hubbard-like models (several, not one) that do conform with the above. They are dismissed~\cite{comment} as having a ``particular symmetry''. Let me give another spin-1/2 ladder~\cite{foot} example ($q=4$) that has hopefully less ``symmetry''. The Floquet dynamics uses two gates, one is a density correlated hopping on the lower leg $U_{\rm zzXX}=\exp{(-\ii \frac{\pi}{4} \sz{1}\sz{2}(\tx{1}\tx{2}+\ty{1}\ty{2}))}$ (or an analogous $\tilde{U}_{\rm zzXX}$ with the correlated hopping on the upper leg, see~\cite{my} for notation), the other a non-conserving gate $U_{\rm G}$ on the lower leg used in Ref.~\cite{my}. At each step (there are $2(L-1)$ per unit of time; $L$ is the number of rungs) we apply a unitary on a randomly selected plaquete, and compare two models: (i) apply either $U_{\rm zzXX}$ or $\tilde{U}_{\rm zzXX}$, (ii) apply either $U_{\rm zzXX}$ or $\tilde{U}_{\rm zzXX}U_{\rm G}$. The model (i) conserves the total spin on the upper and the lower leg (two diffusive DOF), while (ii) conserves spin only on the upper leg (one diffusive DOF). In Fig.~\ref{fig:plotzzXX} we see that, in line with Ref.~\cite{my}, the model (ii) displays ballistic $S_2 \sim t$ rather than diffusive growth. In fact, even the model (i) can display ballistic growth if the bipartite cut is parallel to the direction of diffusive spreading!

{\em Random circuit with $q=3$.--} 
Ref.~\cite{comment} shows numerical data for a particular random circuit suggesting $S_2 \sim \sqrt{t}$ for $q=3$. The only evidence is numerically calculated $dS_2/dt$, which is compatible with $\sim 1/\sqrt{t}$ in a window $dS_2/dt \in [0.3,0.4]$ (or $t \in [25,50]$). Considering a fitting of a power-law in a tiny window I would not quite call that a ``direct refutation''~\cite{comment}, \new{however, let us nevertheless assume $S_2 \sim \sqrt{t}$ is correct}.

Such observation could possibly be explained by special properties of random circuits. In particular, it has been known for some time that random circuits with 2-qubit gates and the full 1-site invariance (Haar random single-qubit gates) can be mapped to Markovian chains on a reduced space~\cite{exact,Viola10} -- due to the invariance the operator space for the average dynamics has dimension $2$ instead of the full $q^2$ -- on top of that, this average dynamics can be for $q=2$ and either random or specific fixed 2-qubit gates described by integrable spin chains~\cite{exact}. The average dynamics of such random circuits is therefore doubly special -- it lives on a reduced space on which it is described by an integrable model. Similar dimensionality reduction could be at play also in random circuits with conserved DOFs~\cite{prl19}. If true, this could explain diffusive growth of $S_2$; even-though one seemingly has $q=3$, the average dynamic is effectively that of a two-level model.
 
{\em Proofs of diffusive growth.--}
In Ref.~\cite{comment} they argue that $S_2 \sim \sqrt{t}$ is expected in diffusive systems due to Ref.~\cite{huang20}, which shows that a so-called ``frozen regions'' (FR)~\cite{comment} present a bottleneck to evolution also in $q>2$, eventually resulting in diffusive growth of $S_2$. In line with point (c) one needs to mention that the proof works only under rather specific conditions. The main ingredient (Condition 1, called ``diffusion'' in Ref.~\cite{huang20}, which, however, is not the same as traditional diffusion) is, schematically, a property that if one starts with a state that contains a FR $\ket{0\ldots 0}$, like in a state $\ket{\psi(0)}=\ket{\phi_1}\ket{00\ldots 0}\ket{\phi_2}$, the FR must remain frozen for any $\phi_{1,2}$ up-to times of order $t < m^2/x_0$, where $m$ is the length of the FR and $x_0$ a state-independent constant. It is a short-time property upto which dynamics stays factorized (nontrivial dynamics happens on longer times; to scale $t$ one has to increase $m$). \new{While the FR condition might hold in some diffusive systems, it is, importantly, different and not equal to diffusion. Examples of system with diffusive DOF that violate it are abundant, e.g., all examples in Ref.~\cite{my} and here.} Standard diffusion has been demonstrated in many systems, the FR property on the other hand has not been shown for any with $q>2$. \new{One does not expect that a single 2-level diffusive DOF will in general cause the FR effect, i.e., blocking the whole large Hilbert space.}

{\em Units of time.--}
Ref.~\cite{my} never claimed that its results are inconsistent with Ref.~\cite{prl19}, as point 3. in Ref.~\cite{comment} tries to suggest. Ref.~\cite{comment} also ``complains'' about the chosen units of times, repeating what has already been mentioned twice in Ref.~\cite{my} (caption in Table I. and on p.2).

As explained~\cite{my}, the chosen units of time have no influence on any of the conclusions and are such as to make finite-size analysis -- a must for any serious claims about the asymptotic behavior (point (a)) -- easier. Namely, with chosen units the curves for $S_2$ and different $L$ overlap, facilitating a read-out of the asymptotic power-law exponent. If any other units would be chosen one would have to each time rescale plots of $S_2$ for different $L$'s in order to have an overlapping curves, making analysis cumbersome. While time units of course do influence the crossover time from ballistic to diffusive growth in $d>1$, it has no influence on the value of $S_2$ at which this happens. The result stressed in Ref.~\cite{my} is that for $d>1$ and in the TDL one will always observe the linear growth of $S_2$ at any finite value of $S_2$. The diffusive growth of $S_2$ is in the TDL~\cite{foot4} pushed to infinitely large values of $S_2$.

{\em Meaning of purity ${\rm e}^{S_2}$.--} Everything that the authors of~\cite{comment} say in their point 4. is correct, however, they have crucially omitted that~\cite{my} presents the explanation as ``A non-rigorous intuitive...'', i.e., meant for non-specialist not familiar with the concept. \new{While the statement is not true for special states, it is correct for most states from the Hilbert space (e.g., random states according to the Haar measure all have $S_r= \ln{(N_{\rm A})}-c(r)$, where $c(r)$ is independent of the Hilbert space size $N_{\rm A}$).}


\begin{thebibliography}{99}
\bibitem{foot2} In practice one often has to deal with behavior at finite times (finite time-window), however, in order to be able to claim a given power the window should be reasonably large, e.g., span more than say an order of magnitude.
\bibitem{foot3} Ref.~\cite{comment} on the other hand focuses only on large-spin models. They of course do exist \new{and are important}, however, viewing large $q$ as \new{solely due to a large spin is less generic}. If spin models do show diffusive growth that would be interesting. \new{So far} no hard evidence has been presented in support of that. Numerical results in~\cite{my}, albeit on small systems, are compatible with ballistic growth of $S_2$ also in spin $S=1$ systems.
\bibitem{foot} One can view the two legs as representing spin-up and spin-down fermions.
\bibitem{foot4} Even in finite systems diffusive growth \new{is pushed to large} $S_2 \approx 10^4$ (Fig.2c in~\cite{my}).
\end{thebibliography}
\end{document}